\documentclass[12pt]{article}%
\usepackage{amsmath}
\usepackage{cite}
\usepackage{graphicx}
\usepackage{amsfonts}
\usepackage{amssymb}%
\providecommand{\U}[1]{\protect\rule{.1in}{.1in}}

\csname @addtoreset\endcsname{equation}{section}
\newcommand{\eq}{\begin{equation}}
\newcommand{\feq}{\end{equation}}
\newcommand{\eqn}{\begin{eqnarray}}
\newcommand{\feqn}{\end{eqnarray}}
\newcommand{\arr}{\begin{eqnarray*}}
\newcommand{\farr}{\end{eqnarray*}}

\begin{document}
\begin{titlepage}
\begin{center}\vspace*{-1.0cm}
{\Large{\bf Melvin Space-times in Supergravity}}
\vskip1cm
\vskip 1.3cm
W. A. Sabra
\vskip 1cm
{\small{\it
Physics Department,
American University of Beirut\\ Lebanon  \\}}
\end{center}
\bigskip
\begin{center}
{\bf Abstract}
\end{center}
We consider Melvin-like cosmological and static solutions for the theories of  ${\cal N}=2$, $D=4$ supergravity
coupled to vector multiplets. We analyze the equations of motion and give some explicit solutions
with one scalar and two gauge fields. Generalized Melvin solutions with four charges
are also constructed for an embedding of a truncated ${\cal N}=8$ supergravity theory.
Our results are then extended  to supergravity theories with the scalar manifolds $SL(N, R)/SO(N, R)$. It is shown  that solutions with $N$ charges only exist for $N=8$, $6$ and  $5$ corresponding to theories with
space-time dimensions $D=4$, $5$ and $7$.
\end{titlepage}

\section{Introduction}

Recently an active area of research has been the study of supersymmetric
gravitational backgrounds in supergravity theories with various space-times
dimensions and signatures \cite{report}. Our present work will only focus on
non-supersymmetric time-dependent and static solutions in some supergravity
theories. Time dependent solutions in string theory and their relevance to
questions in cosmology have been considered by many authors (see for example
\cite{gal, td2, td3, td4, td5, td1} and references therein).

Many years ago, an interesting class of vacuum solutions for Einstein gravity
depending on one variable was constructed by Kasner \cite{kasner}. Related
solutions were also found by several authors \cite{weyl, levi, wilson,
harvey2}. The Kasner metric can be generalized to all space-time signatures
and dimensions \cite{harvey}. The $D$-dimensional Kasner vacuum solution can
take the form%
\begin{equation}
ds^{2}=\epsilon_{0}d\tau^{2}+%
{\displaystyle\sum\limits_{i=1}^{D-1}}
\epsilon_{i}\tau^{2a_{i}}dx_{i}^{2} \label{kas}%
\end{equation}
where $\epsilon_{0},\epsilon_{i}=\pm1$. The so-called Kasner exponents
satisfy
\begin{equation}%
{\displaystyle\sum\limits_{i=1}^{D-1}}
a_{i}=%
{\displaystyle\sum\limits_{i=1}^{D-1}}
a_{i}^{2}=1\text{ }. \label{kgcon}%
\end{equation}
Four-dimensional charged Kasner universes of the form (\ref{kas}) with
fixed\ exponents, as solutions admitting Killing spinors, were considered in
\cite{phantom}. Generalized Melvin-fluxtubes, domain walls and cosmologies for
Einstein-Maxwell-dilaton theories and the correspondence among them were
explored in \cite{kt}. These generalized solutions were obtained by applying a
solution generating technique to seed Levi-Civita and Kasner space-times. The
results of \cite{kt} were later extended to gravitational theories with a
dilaton and an arbitrary rank antisymmetric tensor in \cite{fb}. There,
explicit solutions were constructed for the various $D=11$ supergravity and
type II supergravity theories constructed by Hull \cite{Hull}. Moreover,
Melvin space-times were studied in \cite{s2} for the ungauged five-dimensional
$\mathcal{N}=2$ supergravity theory. The equations of the very special
geometry underlying the structure of the five dimensional theories turned out
to be extremely useful in the analysis of the equations of motion and in the
construction of the solutions.

Our present work shall mainly extend the results of \cite{kt, fb, s2} and deal
with non-supersymmetric Melvin-like cosmological and static solutions of the
theories of ungauged four-dimensional $\mathcal{N}=2$ supergravity theories
coupled to vector multiplets in arbitrary space-time signature
\cite{superbook}. In these theories, the geometry of the scalar fields is a
direct product of the geometry of vector multiplets scalars and that of the
hypermultiplets scalars. In the present work, the hypermultiplets are ignored
and only the scalars of vector multiplets are kept. For the detailed study of
the extension of special geometry to theories with Euclidean and neutral
signatures, we refer the reader to \cite{m1,m2,m3,m4,m5}. The vector multiplet
sector of four-dimensional supergravity in arbitrary space-time signature has
been considered in \cite{m6} via the reduction of Hull's eleven-dimensional
supergravity theories on Calabi-Yau threefolds, followed by a reduction on
spacelike and timelike circles. Moreover, four-dimensional $\mathcal{N}=2$
supergravity coupled to vector and hypermultiplets in signatures $(0,4)$,
$(1,3)$ and $(2,2)$ were obtained via the compactification of type-II string
theories with signatures $(0,10)$, $(1,9)$ and $(2,8)$ on Calabi-Yau
threefolds \cite{mg}.

Our work is planned as follows. In the next section we briefly present some of
the basic properties of the ungauged four-dimensional $\mathcal{N}=2$
supergravity theories and their equations of motion for the metric, gauge and
scalar fields. In section three, we perform an analysis of the equations of
motion and derive our solutions. Section four contains some explicit solutions
for two inequivalent Lorentzian supergravity models where the scalar manifold
is given by $SL(2,R)/SO(2)$ and for an embedding of a truncated $\mathcal{N}%
=8$ supergravity. In section five, we present two classes of solutions for
supergravity theories where the scalars lie in the coset $SL(N,R)/SO(N,R)$. It
is demonstrated that solutions with $N$ charges exist only for $N=8,$ $6$ and
$5$ corresponding to space-time dimensions $D=4,$ $5$ and $7$. Our results are
summarized in section six.

\section{4D Supergravity}

The Lagrangian of the general theory of ungauged $\mathcal{N}=2$, $D=4$
supergravity theories can be given by
\begin{equation}
\mathcal{L}_{4}=\sqrt{|g|}\left[  R-2Q_{IJ}\partial_{\mu}X^{I}\partial^{\mu
}{\bar{X}}^{J}-\frac{\alpha}{2}\left(  \operatorname{Im}\mathcal{N}%
_{IJ}\mathcal{F}^{I}\cdot\mathcal{F}^{J}+\operatorname{Re}\mathcal{N}%
_{IJ}\mathcal{F}^{I}\cdot\mathcal{\tilde{F}}^{J}\right)  \;\right]  \text{ }.
\label{Action}%
\end{equation}
The theory has $n+1$ gauge fields $A^{I},(\mathcal{F}^{I}=d\mathcal{A}^{I})$
and $X^{I}$ are functions of $n$ complex scalar fields $z^{a}.$ The details of
the Lorentzian four-dimensional supergravity theories and their formulation in
terms of special geometry can be found in \cite{superbook}. Our analysis is
not restricted to Lorentzian theories and is valid for all space-time
signatures. The parameter $\alpha$ takes the values $\pm1$. Roughly speaking,
the theories of Euclidean and $(2,2)$ signature can be obtained by replacing
the complex structure with a paracomplex structure \cite{m1,m2,m3,m4,m5}. To
use a unified description, we define $i_{\epsilon}$ which satisfies
$i_{\epsilon}^{2}=\epsilon$ and $\bar{\imath}_{\epsilon}=-i_{\epsilon}$. Here
$\epsilon=1$ for the theories with Euclidean and neutral signature and
$\epsilon=-1$ for the Lorentzian theory. We note the relation
\begin{equation}
Q_{IJ}\partial_{\mu}X^{I}\partial^{\mu}{\bar{X}}^{J}=g_{a\bar{b}}\partial
_{\mu}z^{a}\partial^{\mu}\bar{z}^{b}%
\end{equation}
where$\ g_{a\bar{b}}=\partial_{a}\partial_{\bar{b}}K$ is the K\"{a}hler metric
and $K$ is the K\"{a}hler potential of the supergravity theory.

In a formulation of special geometry, one relates the coordinates $X^{I}$ to
the covariantly holomorphic sections
\begin{align}
V  &  =\left(
\begin{array}
[c]{c}%
L^{I}\\
M_{I}%
\end{array}
\right)  ,\text{ \ \ \ }I=0,...,n\nonumber\\
\mathcal{D}_{\bar{a}}V  &  =\left(  \partial_{\bar{a}}-\frac{1}{2}%
\partial_{\bar{a}}K\right)  V=0,
\end{align}
obeying the constraint
\begin{equation}
i_{\epsilon}\langle V,\bar{V}\rangle=i_{\epsilon}\left(  \bar{L}^{I}%
M_{I}-L^{I}\bar{M}_{I}\right)  =1,
\end{equation}
by
\begin{equation}
\Omega=e^{-K/2}V=\left(
\begin{array}
[c]{c}%
X^{I}\\
F_{I}%
\end{array}
\right)  ,\text{ \ \ \ \ }\partial_{\bar{a}}\Omega=0\text{ }.
\end{equation}
The K\"{a}hler potential is given by
\begin{equation}
e^{-K}=i_{\epsilon}\left(  \bar{X}^{I}F_{I}-X^{I}\bar{F}_{I}\right)  \text{ }.
\end{equation}
We also have the relations
\begin{align}
M_{I}  &  =\mathcal{N}_{IJ}L^{J},\text{ \ \ \ \ \ }\mathcal{D}_{a}%
M_{I}=\mathcal{\bar{N}}_{IJ}\mathcal{D}_{a}L^{I},\nonumber\\
\mathcal{D}_{a}  &  =\left(  \partial_{a}+\frac{1}{2}\partial_{a}K\right)
\text{ }.
\end{align}
In cases where the $\mathcal{N}=2$ supergravity models can be described in
terms of a holomorphic homogeneous prepotential $F=F(X^{I})$ of degree two, we
have $F_{I}=\frac{\partial F}{\partial X^{I}}$, $F_{IJ}=\frac{\partial
F}{\partial X^{I}\partial X^{J}}$, etc. Here we list the following useful relations%

\begin{align}
F  &  ={\frac{1}{2}}F_{I}X^{I},\text{ \ \ \ }F_{I}=F_{IJ}X^{J},\nonumber\\
X^{I}F_{IJK}  &  =0,\text{ \ \ \ \ \ }F_{I}\partial_{\mu}X^{I}=X^{I}%
\partial_{\mu}F_{I}\text{ }. \label{fun2}%
\end{align}
The scalar and gauge couplings appearing in the Lagrangian are given by
\begin{align}
\mathcal{N}_{IJ}  &  =\bar{F}_{IJ}-\epsilon{\frac{i_{\epsilon}}{(XNX)}%
}(NX)_{I}(NX)_{J},\nonumber\\
Q_{IJ}  &  =e^{K}N_{IJ}+e^{2K}(N{\bar{X}})_{I}(NX)_{J}, \label{moh}%
\end{align}
where
\begin{equation}
N_{IJ}=i_{\epsilon}\left(  \bar{F}_{IJ}-F_{IJ}\right)  , \label{moh1}%
\end{equation}
and with the notation $(NX)_{I}=N_{IJ}X^{J}$, $(N{\bar{X}})_{I}=N_{IJ}{\bar
{X}}^{J}$, $XNX=X^{I}X^{J}N_{IJ}$.

The gauge fields equations of motion derived from (\ref{Action}) are given by%
\begin{equation}
\partial_{\mu}\left[  \sqrt{|\det g|}\left(  \operatorname{Im}\mathcal{N}%
_{IJ}\mathcal{F}^{J\mu\nu}{}-\frac{1}{2}(\operatorname{Re}\mathcal{N}%
_{IJ})\epsilon^{\mu\nu\rho\sigma}\mathcal{F}{^{J}}_{\rho\sigma}\right)
\right]  =0\text{ }. \label{max}%
\end{equation}
The field equations of the scalars $z^{a}$ and $\bar{z}^{a}$ are%
\begin{align}
&  \left(  \frac{1}{\sqrt{|\det g|}}\partial_{\mu}\left(  \sqrt{|\det
g|}Q_{IJ}\partial^{\mu}{\bar{X}}^{J}\right)  -\left(  \partial_{L}%
Q_{IJ}\right)  \partial_{\mu}X^{I}\partial^{\mu}{\bar{X}}^{J}\right)
\partial_{a}{\bar{X}}^{L}\nonumber\\
&  -\alpha\left(  \frac{1}{4}\partial_{L}(\operatorname{Im}\mathcal{N}%
_{IJ})\mathcal{F}{^{I}}_{\mu\nu}\mathcal{F}^{J\ \mu\nu}+\frac{1}{8\sqrt{|\det
g|}}\partial_{L}(\operatorname{Re}\mathcal{N}_{IJ})\epsilon^{\mu\nu\rho\sigma
}\mathcal{F}{^{I}}_{\mu\nu}\mathcal{F}{^{J}}_{\rho\sigma}\right)  \partial
_{a}X^{L}\nonumber\\
&  =0\ \label{scalareq1}%
\end{align}
and%

\begin{align}
&  \left(  \frac{1}{\sqrt{|\det g|}}\partial_{\mu}\left(  \sqrt{|\det
g|}Q_{LJ}\partial^{\mu}{X}^{J}\right)  -(\partial_{\bar{L}}Q_{IJ}%
)\partial_{\mu}X^{I}\partial^{\mu}{\bar{X}}^{J}\right)  \partial_{a}%
X^{L}\nonumber\\
&  -\alpha\left(  \frac{1}{4}\partial_{\bar{L}}(\operatorname{Im}%
\mathcal{N}_{IJ})\mathcal{F}{^{I}}_{\mu\nu}\mathcal{F}^{J\ \mu\nu}+\frac
{1}{8\sqrt{|\det g|}}\partial_{\bar{L}}(\operatorname{Re}\mathcal{N}%
_{IJ})\epsilon^{\mu\nu\rho\sigma}\mathcal{F}{^{I}}_{\mu\nu}\mathcal{F}{^{J}%
}_{\rho\sigma}\right)  \partial_{a}{\bar{X}}^{L}\nonumber\\
&  =0\ .
\end{align}
The Einstein equations of motion are given by
\begin{equation}
R_{\mu\nu}=2g_{a\bar{b}}\partial_{\mu}z^{a}\partial_{\nu}\bar{z}^{b}%
+\alpha\operatorname{Im}\mathcal{N}_{IJ}\left(  \mathcal{F}^{I}{}_{\mu\lambda
}\mathcal{F}^{J}{}_{\nu}{}^{\lambda}-{\frac{1}{4}}g_{\mu\nu}\mathcal{F}^{I}%
{}_{\alpha\beta}\mathcal{F}^{J}{}^{\alpha\beta}\right)  \text{ }. \label{ein}%
\end{equation}

\section{Solutions}

We consider solutions of the form
\begin{equation}
ds^{2}=e^{2U}\left(  \epsilon_{0}d\tau^{2}+\epsilon_{1}\tau^{2a}%
dx^{2}+\epsilon_{2}\tau^{2b}dy^{2}\right)  +e^{-2U}\epsilon_{3}\tau^{2c}dz^{2}
\label{gen}%
\end{equation}
where $U$ is a function of $\tau.$ We obtain from (\ref{max}) for a
non-vanishing $\mathcal{F}_{\tau z}^{I}{},$ the solutions
\begin{equation}
\operatorname{Im}\mathcal{N}_{IJ}\mathcal{F}^{J}{}_{\tau z}=q_{I}\tau
^{2c-1}e^{-2U} \label{mas}%
\end{equation}
where $q_{I}$ are constants. The non-vanishing components of the Ricci tensor
for the metric (\ref{gen}) are given by%

\begin{align}
R_{\tau\tau}  &  =-\ddot{U}-\left(  1-4c\right)  \frac{\dot{U}}{\tau}-2\dot
{U}^{2},\nonumber\\
R_{xx}  &  =-\epsilon_{0}\epsilon_{1}\tau^{2a}\left(  \ddot{U}+\frac{\dot{U}%
}{\tau}\right)  ,\nonumber\\
R_{yy}  &  =-\epsilon_{0}\epsilon_{2}\tau^{2b}\left(  \ddot{U}+\frac{\dot{U}%
}{\tau}\right)  ,\nonumber\\
R_{zz}  &  =\epsilon_{0}\epsilon_{3}e^{-4U}\tau^{2c}\left(  \ddot{U}%
+\frac{\dot{U}}{\tau}\right)  \text{ }. \label{genera}%
\end{align}
Using (\ref{mas}) and (\ref{genera}) and the Einstein equations of motion
(\ref{ein}), we obtain, for real scalars, the two equations%

\begin{align}
\ddot{U}+\left(  1-2c\right)  \frac{\dot{U}}{\tau}+\dot{U}^{2}  &
=-Q_{IJ}\dot{X}^{J}\dot{X}^{I},\label{e}\\
\ddot{U}+\frac{\dot{U}}{\tau}  &  =\frac{\alpha}{2}\epsilon_{3}e^{-2U}%
\operatorname{Im}\mathcal{N}^{IJ}q_{I}q_{J}\tau^{2c-2}\text{ }. \label{eqt}%
\end{align}
We now employ the relations of special geometry in the analysis of the
equations (\ref{e}) and (\ref{eqt}). For real $X^{I}$, the prepotential $F$
and all its derivatives are purely imaginary. In this case, we obtain from
(\ref{moh}) the following relations
\begin{align}
Q_{IJ}  &  =\frac{1}{2F}\left(  \frac{F_{I}F_{J}}{2F}-F_{IJ}\right)
,\nonumber\\
\mathcal{N}_{IJ}  &  ={\frac{F_{I}F_{J}}{F}}-F_{IJ}\text{ }. \label{re}%
\end{align}
Using (\ref{fun2}) and (\ref{re}), we obtain the relation
\begin{equation}
Q_{IJ}\dot{X}^{J}\dot{X}^{I}=\frac{1}{2F}\left(  \frac{\dot{F}^{2}}{2F}%
-\ddot{F}+X^{I}\ddot{F}_{I}\right)  \text{ }. \label{n}%
\end{equation}
As an ansatz for our solution we take%

\begin{equation}
e^{2U}=4i_{\epsilon}F, \label{an}%
\end{equation}
then we obtain from (\ref{n}) and special geometry the relations
\begin{align}
Q_{IJ}\dot{X}^{J}\dot{X}^{I}  &  =-\ddot{U}\text{ }-\dot{U}^{2}+2e^{-2U}%
\operatorname{Im}\mathcal{N}^{IJ}F_{J}\ddot{F}_{I},\nonumber\\
\dot{U}\text{ }  &  =2e^{-2U}\operatorname{Im}\mathcal{N}^{IJ}\dot{F}_{I}%
F_{J},\nonumber\\
\ddot{U}  &  =2e^{-2U}{\operatorname{Im}\mathcal{N}^{IJ}}\left(  F_{J}\ddot
{F}_{I}-{\dot{F}_{J}}\dot{F}_{I}\right)  \text{ }.
\end{align}
Using these relations in (\ref{e}) and (\ref{eqt}), we finally obtain%

\begin{align}
\left(  1-2c\right)  \frac{X^{I}\dot{F}_{I}}{\tau}+X^{I}\ddot{F}_{I}  &
=0,\nonumber\\
2\operatorname{Im}\mathcal{N}^{IJ}\left(  {\dot{F}_{J}}\dot{F}_{I}-F_{J}%
\ddot{F}_{I}-\frac{F_{J}\dot{F}_{I}}{\tau}\right)  +{\frac{\alpha}{2}}%
\epsilon_{3}\operatorname{Im}\mathcal{N}^{IJ}q_{I}q_{J}\tau^{2c-2}  &
=0\text{ }. \label{dah}%
\end{align}
The first equation in (\ref{dah}) can be solved by
\begin{equation}
F_{I}=\frac{1}{2}\epsilon i_{\epsilon}\left(  A_{I}+B_{I}\tau^{2c}\right)
\label{so}%
\end{equation}
with constants $A_{I}$ and $B_{I.}$ The second equation then reduces to the
algebraic condition%

\begin{equation}
\operatorname{Im}\mathcal{N}^{IJ}\left(  2c^{2}\epsilon A_{J}B_{I}%
-\frac{\alpha}{2}\epsilon_{3}q_{I}q_{J}\right)  =0\text{ }. \label{con1}%
\end{equation}
It remains to analyze the scalar equations of motion. After some calculation
employing the equations of special geometry, the scalar equations of motion
reduce to the algebraic condition
\begin{equation}
\partial_{a}\operatorname{Im}\mathcal{N}^{IJ}\left(  2c^{2}\epsilon A_{J}%
B_{I}-\frac{\alpha}{2}\epsilon_{3}q_{I}q_{J}\right)  =0\text{ }. \label{con2}%
\end{equation}

A dual solution can be obtained where we have a non-vanishing $\mathcal{F}%
_{xy}^{I}{}=p^{I}.$ In this case, $X^{I}$ are imaginary. The analysis of
Einstein equations of motion then gives%

\begin{align}
\ddot{U}+\left(  1-2c\right)  \frac{\dot{U}}{\tau}+\dot{U}^{2}  &  =Q_{IJ}%
\dot{X}^{I}{\dot{X}}^{J},\nonumber\\
\ddot{U}+\frac{\dot{U}}{\tau}  &  =-\frac{1}{2}\epsilon_{0}\epsilon
_{1}\epsilon_{2}\alpha e^{-2U}\tau^{-2\left(  a+b\right)  }\operatorname{Im}%
\mathcal{N}_{IJ}p^{I}p^{J}\text{ }. \label{ee}%
\end{align}
Again we take the solution%

\begin{equation}
e^{2U}=-4i_{\epsilon}F\text{ }.
\end{equation}
Special geometry relations give the following equations%

\begin{align}
\dot{U}  &  =-2i_{\epsilon}\dot{X}^{I}F_{I}e^{-2U}\nonumber\\
\ddot{U}  &  =-2i_{\epsilon}\left(  \ddot{X}^{I}F_{I}-\mathcal{N}{_{IJ}}%
\dot{X}^{I}{\dot{X}^{J}}\right)  e^{-2U}\text{ }. \label{us}%
\end{align}
Note that in this case the relations (\ref{moh}) for imaginary $X^{I}$ imply
\begin{equation}
Q_{IJ}\dot{X}^{J}\dot{X}^{I}=-\frac{1}{2F}\mathcal{N}_{IJ}\dot{X}^{I}\dot
{X}^{J}+\dot{U}^{2}\text{ }. \label{us1}%
\end{equation}
Consequently, using (\ref{us}) and (\ref{us1}), the equations (\ref{ee})
reduce to%

\begin{align}
\frac{F_{I}}{F}\left[  \ddot{X}^{I}+\left(  1-2c\right)  \frac{\dot{X}^{I}%
}{\tau}\right]   &  =0,\nonumber\\
\operatorname{Im}\mathcal{N}_{IJ}\left(  \ddot{X}^{I}X^{J}-\dot{X}^{I}{\dot
{X}^{J}}+\frac{X^{J}\dot{X}^{I}}{\tau}\right)   &  =\frac{1}{4}\epsilon
_{0}\epsilon_{1}\epsilon_{2}\epsilon\alpha\operatorname{Im}\mathcal{N}%
_{IJ}\tau^{-2\left(  a+b\right)  }p^{I}p^{J}\text{ }. \label{ma}%
\end{align}
The first equation in (\ref{ma}) admits the solution
\begin{equation}
X^{I}=\frac{i_{\epsilon}}{2}\left(  A^{I}+B^{I}\tau^{2c}\right)
\end{equation}
which upon plugging in the second equation of (\ref{ma}) gives the algebraic condition%

\begin{equation}
\operatorname{Im}\mathcal{N}_{IJ}\left(  c^{2}B^{I}A^{J}-\frac{1}{4}%
\epsilon_{0}\epsilon_{1}\epsilon_{2}\alpha p^{I}p^{J}\right)  =0\text{ }.
\end{equation}
Again the analysis of the scalar equation gives, after some calculation
involving special geometry relations, the algebraic condition
\begin{equation}
\partial_{a}\operatorname{Im}\mathcal{N}_{IJ}\left(  c^{2}B^{I}A^{J}-\frac
{1}{4}\epsilon_{0}\epsilon_{1}\epsilon_{2}\alpha p^{I}p^{J}\right)  =0\text{
}.
\end{equation}

\section{Examples}

Consider the $\mathcal{N}=2$ supergravity model with a Lorentzian signature
and with the prepotential $F=-iX^{0}X^{1}.$ This corresponds to a model where
the scalar manifold is given by $SL(2,R)/SO(2).$ For this model we obtain from
(\ref{so}) the solution
\begin{equation}
X^{1}=\frac{1}{2}\left(  1+B_{0}t^{2c}\right)  ,\text{ \ \ }X^{0}=\frac{1}%
{2}\left(  1+B_{1}t^{2c}\right)  \text{ }.
\end{equation}
Using (\ref{an}) and the algebraic conditions (\ref{con1}) and (\ref{con2})
(with $\epsilon=\alpha=-\epsilon_{3}=-1),$ we finally arrive at the metric
\begin{equation}
ds^{2}=\left(  1+\frac{q_{0}^{2}}{4c^{2}}t^{2c}\right)  \left(  1+\frac
{q_{1}^{2}}{4c^{2}}t^{2c}\right)  \left(  -dt^{2}+t^{2a}dx^{2}+t^{2b}%
dy^{2}\right)  +\frac{t^{2c}}{\left(  1+\frac{q_{0}^{2}}{4c^{2}}t^{2c}\right)
\left(  1+\frac{q_{1}^{2}}{4c^{2}}t^{2c}\right)  }dz^{2}\text{ }. \label{fs}%
\end{equation}
Using (\ref{mas}) and the second equation in (\ref{re}), we obtain for the
gauge fields
\begin{equation}
\mathcal{F}_{tz}^{0}{}=-\frac{q_{0}t^{2c-1}}{\left(  1+\frac{q_{0}^{2}}%
{4c^{2}}t^{2c}\right)  ^{2}},\text{\ \ }\mathcal{F}_{tz}^{1}=-\frac
{q_{1}t^{2c-1}}{\left(  1+\frac{q_{1}^{2}}{4c^{2}}t^{2c}\right)  ^{2}}\text{
}.
\end{equation}
The dual solution with $\mathcal{F}_{xy}^{0}{}=p^{0}$ and $\mathcal{F}%
_{xy}^{1}{}=p^{1}$ has the scalar fields given by
\begin{equation}
X^{0}=\frac{i}{2}\left(  1+\frac{\left(  p^{0}\right)  ^{2}}{4c^{2}}%
t^{2c}\right)  ,\text{ \ \ \ \ }X^{1}=\frac{i}{2}\left(  1+\frac{\left(
p^{1}\right)  ^{2}}{4c^{2}}t^{2c}\right)  ,
\end{equation}
and the metric as in (\ref{fs}) with $q_{0}$ and $q_{1}$ replaced by $p^{0}$
and $p^{1}.$

These solutions can be referred to as generalized Melvin cosmologies
\cite{kt}. Similarly we can also construct Melvin domain wall solutions%

\begin{equation}
ds^{2}=\left(  1+\frac{q_{0}^{2}}{4c^{2}}r^{2c}\right)  \left(  1+\frac
{q_{1}^{2}}{4c^{2}}r^{2c}\right)  \left(  dr^{2}-r^{2a}dt^{2}+r^{2b}%
dy^{2}\right)  +\frac{r^{2c}}{\left(  1+\frac{q_{0}^{2}}{4c^{2}}r^{2c}\right)
\left(  1+\frac{q_{1}^{2}}{4c^{2}}r^{2c}\right)  }dz^{2}\text{ }. \label{f}%
\end{equation}

Taking $z$ as an angle coordinate gives Melvin fluxtubes with two charges. The
original Melvin fluxtube \cite{melvin} can be obtained using our formalism
with $F=-i\left(  X^{0}\right)  ^{2}$ and the exponents $a=b=0$ and $c=1.$

One can also consider solutions of the so-called minimal coupling model
described by the prepotential $F=-i\sqrt{X^{0}\left(  X^{1}\right)  ^{3}}$ and
scalar manifold $SL(2,R)/SO(2)$. For this model we have%

\begin{equation}
F_{0}=\frac{F}{2X^{0}},\text{ \ \ \ }F_{1}=\frac{3F}{2X^{1}}%
\end{equation}
and%

\begin{equation}
\mathcal{N}_{00}={\frac{F}{2\left(  X^{0}\right)  ^{2}},}\text{ \ \ \ \ \ \ }%
\mathcal{N}_{11}={\frac{3F}{2\left(  X^{1}\right)  ^{2}},}\text{
\ \ \ }\mathcal{N}_{01}={0}%
\end{equation}
and explicit solutions can be obtained using (\ref{mas}), (\ref{an}),
(\ref{so}), (\ref{con1}) and (\ref{con2}).

As another example we consider solutions of $\mathcal{N}=8$, $SO(8)$
supergravity \cite{N=8} by focusing on the $U(1)^{4}$ Cartan subgroup. Anti-de
Sitter black holes solutions of the gauged version of this theory were
considered in \cite{dlgauged, anti}. The resulting model can be embedded in an
$\mathcal{N}=2$ supergravity model with the following prepotential
\begin{equation}
F=-i\sqrt{X^{1}X^{2}X^{3}X^{4}}\text{ }. \label{pre}%
\end{equation}
Note that the previous two models considered are also consistent truncation of
$\mathcal{N}=8$, $SO(8)$ supergravity.

Using (\ref{re}) and (\ref{pre}) we obtain%

\begin{equation}
F_{I}=\frac{F}{2X^{I}},\text{ \ \ }\mathcal{N}_{IJ}=\frac{F}{4\left(
X^{I}\right)  ^{2}}\delta_{IJ}\text{ }.
\end{equation}
Using our analysis we obtain the generalized Melvin cosmological solutions
\begin{equation}
ds^{2}=e^{2U}\left(  -dt^{2}+t^{2a}dx^{2}+t^{2b}dy^{2}\right)  +e^{-2U}%
t^{2c}dz^{2}\text{ }%
\end{equation}
with%
\begin{equation}
e^{2U}=4\sqrt{H_{1}H_{2}H_{3}H_{4}}\text{ },\text{ \ \ }H_{I}=\left(
1+\frac{q_{I}^{2}}{4c^{2}}t^{2c}\right)  \text{ }.
\end{equation}
The scalars and gauge fields are given by%

\begin{equation}
X^{I}=\frac{\sqrt{H_{1}H_{2}H_{3}H_{4}}}{H_{I}},\text{ \ \ \ \ }%
\mathcal{F}^{I}{}_{\tau z}=-\frac{q_{I}t^{2c-1}}{\left(  H_{I}\right)  ^{2}%
}\text{ }.
\end{equation}
Similarly one can also obtain generalized Melvin domain wall and fluxtube solutions.

\section{$SL(N,R)/SO(N,R)$ coset models}

In this section we consider solutions to $D$-dimensional supergravity theories
with scalar fields parameterizing the space $SL(N,R)/SO(N,R)$. These theories
can be described by the following Lagrangian%

\begin{equation}
e^{-1}\,\mathcal{L}_{D}=R-{{\frac{{1}}{{2}}}}(\partial\vec{\varphi}%
)^{2}-{\frac{1}{4}}G_{IJ}\mathcal{F}^{I}\cdot\mathcal{F}^{J}\text{ }.
\label{ddlag}%
\end{equation}
The gauge kinetic term metric given by
\begin{equation}
G_{IJ}={\frac{1}{(X_{I})^{2}}}\delta_{IJ}\text{ }.
\end{equation}
The scalars are described by $X_{I}$ subject to the condition
\begin{equation}
\prod_{I=1}^{N}X_{I}=1\, \label{prod}%
\end{equation}
and are related to the $(N-1)$ independent dilatonic scalars $\vec{\varphi}$
by
\begin{equation}
X_{I}=e^{-{\frac{1}{2}}\vec{b}_{I}\cdot\vec{\varphi}}\text{ } \label{ddxdef}%
\end{equation}
where $\vec{b}_{I}$ are the weight vectors of the fundamental representation
of $SL(N,R)$, satisfying
\begin{equation}
\vec{b}_{I}\cdot\vec{b}_{J}=8\delta_{IJ}-{\frac{8}{N}}\,,\qquad\sum_{I}\vec
{b}_{I}=0\text{ }. \label{dotprod}%
\end{equation}
The scalars $\vec{\varphi}$ can be described in terms of $X_{I}$ by
\begin{equation}
\vec{\varphi}=-{{\frac{{1}}{{4}}}}\sum_{I}\vec{b}_{I}\,\log X_{I}\,\text{\ }.
\label{quake}%
\end{equation}
Domain wall and charged time-dependent solutions for the gauged versions of
these theories were considered in \cite{Cv} and \cite{gutsabra}.

We start by considering solutions of the form
\begin{equation}
ds^{2}=e^{2U\left(  \tau\right)  }\left(  \epsilon_{0}d\tau^{2}+\epsilon
_{1}\tau^{2a_{1}}dx_{1}^{2}+\epsilon_{2}\tau^{2a_{2}}dx_{2}^{2}\right)
+e^{2V\left(  \tau\right)  }\sum_{k=3}^{D-1}\epsilon_{k}\tau^{2a_{k}}%
dx_{k}^{2}%
\end{equation}
with
\begin{equation}
V=-\frac{1}{D-3}U\text{ }.
\end{equation}
For these solutions, the gauge fields two-form is given by
\begin{equation}
F^{I}=P^{I}dx_{1}\wedge dx_{2}%
\end{equation}
where $P^{I}$ are constants. The analysis of the equations of motion derived
from (\ref{ddlag}) gives the solution
\begin{equation}
X^{I}=e^{-U}\left(  1+B^{I}\tau^{2l}\right)  ,\text{ \ \ \ \ \ \ }e^{NU}%
=\prod_{I=1}^{N}\left(  1+B^{I}\tau^{2l}\right)
\end{equation}
with
\begin{equation}
l=\sum_{k=3}^{D-1}a_{k}%
\end{equation}
provided the conditions
\begin{equation}
{\frac{N\left(  D-3\right)  }{4(D-2)}}=1,\text{ \ \ \ }B^{I}=-\frac{1}{2l^{2}%
}\epsilon_{0}\epsilon_{1}\epsilon_{2}\left(  P^{I}\right)  ^{2} \label{nice}%
\end{equation}
are satisfied. The analysis of the scalar equations reveals no further
conditions. The condition (\ref{nice}) was also obtained in the study of
domain wall solutions and S-branes \cite{Cv, gutsabra}. As both space-time
dimensions and $N$ must be integers, it is evident that our solutions are only
valid for the space-time dimensions $D=4,$ $5$ and $7$ corresponding to the
cases $N=8,$ $6$ and $5$.

A second class of solutions, with the condition (\ref{nice}), can be obtained
with the metric%

\begin{equation}
ds^{2}=e^{2U}\left(  \epsilon_{0}d\tau^{2}+\sum_{i=1}^{D-2}\epsilon_{i}%
\tau^{2a_{i}}dx_{i}^{2}\right)  +e^{2(3-D)U}\epsilon_{D-1}\tau^{2a_{D-1}%
}dw^{2},
\end{equation}
where%
\begin{equation}
e^{N(D-3)U}=\text{\ }\prod_{I=1}^{N}\left(  1+B^{I}\tau^{2(a_{D-1})}\right)
,\text{ \ \ \ \ \ }B^{I}=\frac{1}{2\left(  a_{D-1}\right)  ^{2}}\epsilon
_{D-1}\left(  q_{I}\right)  ^{2}%
\end{equation}
and
\begin{equation}
X^{I}=\frac{e^{(D-3)U}}{\left(  1+B^{I}\tau^{2(a_{D-1})}\right)  },\text{
\ \ \ \ \ }{G}_{IJ}F_{\tau w}^{J}=q_{I}e^{2(3-D)U}\tau^{\left(  2a_{D-1}%
-1\right)  }\text{ }.
\end{equation}

\section{Summary}

We have considered solutions depending on one variable for the theories of
four-dimensional $\mathcal{N}=2$ supergravity theories with vector multiplets.
Depending on the signature of the theory, our charged solutions can describe
time-dependent cosmological or static solutions. These solutions which are
labelled as Melvin space-times can be thought of as charged generalizations of
Kasner spaces.\ We found explicit solutions for specific models in
$\mathcal{N}=2$ supergravity with two charges. Solutions with four charges for
a truncation of $\mathcal{N}=8$ supergravity theory that can be embedded in
$\mathcal{N}=2$ supergravity were also presented. Moreover, solutions with $N$
charges for $N=8,$ $6$ and $5,$ corresponding to supergravity theories with
space-time dimensions $D=4,$ $5$ and $7$ and $SL(N,R)/SO(N,R)$ scalar
manifolds were also found.

It is well known that Melvin fluxtubes can be generated from Minkowski
space-time as a seed solution \cite{har}. Generalised Melvin solutions in
Einstein-Maxwell theory were constructed using these techniques in \cite{kt}
with Kasner space being the seed solutions. The generating techniques were
generalized to dilaton gravity in \cite{gendilaton} and to gravity with a
cosmological constant in \cite{astorino}. In our present analysis, we started
with the vacuum Kasner solution as a seed solution and found solutions with
non-trivial scalar and gauge fields through an explicit analysis of the
equations of motion. It would be of interest to study generalized Melvin
solutions in gauged supergravity theories in various dimensions. We hope to
report on this in a future publication.

\bigskip

\textbf{Acknowledgements}: The work is supported in part by the National
Science Foundation under grant number PHY-1620505.

\end{document}